\documentclass[reqno,12pt]{amsart} 
\usepackage[top=1.5in,right=1.125in,left=1.125in,bottom=1.5in]{geometry}
\usepackage{amssymb}
\usepackage{color}
\usepackage{tikz}
\usetikzlibrary{positioning,decorations.pathmorphing}
\usetikzlibrary{positioning,decorations.pathmorphing}
\definecolor{red}{rgb}{0.7,0,0}
\definecolor{grey}{RGB}{112,112,112}
\definecolor{blue}{RGB}{034,113,179}
\usepackage[colorlinks=true,citecolor=blue,linkcolor=red,urlcolor=grey,backref=page]{hyperref}
\newcommand{\koniec}{\begin{flushright}  $\Box $ \end{flushright}}
\newtheorem{theo}{Theorem}[section] 
\newtheorem{prop}[theo]{Proposition}

\theoremstyle{remark}

\newcounter{mnotecount}[section]

\renewcommand{\themnotecount}{\thesection.\arabic{mnotecount}}

\newcommand{\mnote}[1]
{\protect{\stepcounter{mnotecount}}$^{\mbox{\footnotesize
$
\bullet$\themnotecount}}$ \marginpar{
\raggedright\tiny\em
$\!\!\!\!\!\!\,\bullet$\themnotecount: #1} }

\def\vv{\varepsilon}

\newcommand{\R}{\mathbb{R}}

\def\p{\partial}
\def\a{\alpha}
\def\be{\begin{equation}}

\def\g{\gamma}

\def\ee{\end{equation}}

\def\bea{\begin{eqnarray}}
\def\eea{\end{eqnarray}}

\numberwithin{equation}{section}
\begin{document} \date{November 17th, 2021}
\title[The quadric ansatz and Einstein---Weyl]{The quadric ansatz for the $mn$--dispersionless KP equation, and supersymmetric Einstein--Weyl spaces}
\author{Maciej Dunajski}
\address{Department of Applied Mathematics and Theoretical Physics\\ 
University of Cambridge\\ Wilberforce Road, Cambridge CB3 0WA, UK.}
\email{m.dunajski@damtp.cam.ac.uk}
\author{Prim Plansangkate}
\address{Applied Analysis Research Unit, Division of Computational Science\\
Faculty of Science, Prince of Songkla University\\
 Songkhla, 90110 Thailand.}
\email{prim.p@psu.ac.th}
%
\begin{abstract}
We consider two multi--dimensional generalisations of the dispersionless
Kadomtsev-Petviashvili (dKP) equation, both allowing for arbitrary dimensionality, and non--linearity. For one of these generalisations, we characterise all solutions which are constant on a central quadric. The quadric ansatz 
leads to a second order ODE which is equivalent to Painleve I or II for the dKP equation, but fails to pass the Painlev\'e test in higher dimensions. The second generalisation of the dKP equation leads to a class of Einstein--Weyl structures in an arbitrary dimension, which is characterised by the existence of a weighted parallel vector field, together with further holonomy reduction.
We construct and characterise an explicit new
family of Einstein--Weyl spaces belonging to this class, and
depending on one arbitrary function of one variable.
\end{abstract}
\maketitle

\section{Introduction}
Let $u:U\longrightarrow \R$, where $U$ is an open set in $\R^{N}$
with coordinates $x^a=(x, y_1, \dots, y_n, t)$. 
The $mn$--dispersionless Kadomtsev-Petviashvili ($mn$--dKP) equation \cite{MS11,SS16} is given by
\be \label{mndKP}
u_{xt} - (u^m u_x)_x = \triangle u,
\ee
where $\triangle ={\p/\p y_1}^2+\dots+{\p/\p y_n}^2$ and $u_x=\p_x u$, etc. 
We refer the readers to \cite{SS16} for a list of references and applications of equation (\ref{mndKP}) ranging from non-linear optics and acoustics to geometry  for different values of integers $(m, n)$.
\vskip5pt

The aim of this paper is to construct solutions to (\ref{mndKP})
which are constant on central quadrics ${\mathcal Q}\subset\R^{N}$,
i. e. there exists a symmetric $N\times N$ matrix 
${\bf M}={\bf M}(u)$ such that
\be
\label{quadricansatz}
M_{ab}(u)x^ax^b=C, \quad \mbox{where}\quad C=\mbox{const}, \quad\mbox{and}\quad a, b=1, \dots, N=n+2.
\ee  
If $m=n=1$, then (\ref{mndKP}) becomes the standard dKP equation,
and the ODE for the matrix ${\bf M}(u)$ resulting from the quadric ansatz (\ref{quadricansatz}) is either solvable by quadratures, or is equivalent
to Painlev\'e I or Painlev\'e II (this last case being generic) \cite{DT02}. In \S\ref{section2}
we shall perform the analysis of the quadric ansatz 
for (\ref{mndKP}). The non--generic linearisable case leads
to some explicit forms of ${\bf M}(u)$, which we shall present in \S\ref{sec_examples}. The other two cases
lead to 2nd order ODEs generalising Painlev\'e I and II, which however fail to pass the Painlev\'e test (Theorem \ref{theo1}).
Thus, in line with the integrability dogma \cite{ARS80}, we conclude that (\ref{mndKP}) is only integrable
if $m=n=1$ (the dKP case), or $n=0$ (the Riemann equation solvable by the method of characteristics), or $m=0$
(the linear wave equation).
It is known \cite{D01} that the dKP equation characterises a class of Einstein--Weyl metrics in $2+1$ dimensions which admit a parallel weighted vector field. While this correspondence does not extend to 
(\ref{mndKP}) for general $(m, n)$, in \S\ref{sec_ew} we shall demonstrate  that a closely related equation
\be \label{EWeqmndKP}  
u_{xt} - (u u_x)_x + \frac{2(n-1)}{n}u_x^2 = \triangle u, 
\ee
characterises a class of EW structures admitting a parallel null weighted vector field, and  with a further assumption on the holonomy of the Weyl connection. Our main result in \S\ref{sec_ew} is Theorem \ref{theo_EW}
characterising an explicit family of Einstein--Weyl spaces in arbitrary dimension.

\subsection*{Acknowledgements}
MD has been partially supported 
by STFC grants ST/P000681/1, and  ST/T000694/1.
PP is grateful for travel support from the Applied Analysis Research Unit at PSU, and to CIRM in Lumini, where 
some of this research has been carried over during the workshop {\em Twistors meet Loops} in August 2019. 
Both authors thank  Anton Galaev for helpful correspondence.
\section{The quadric ansatz}
\setcounter{equation}{0}

\label{section2}
The quadric ansatz (\ref{quadricansatz}) is applicable to equations of the form
\be 
\label{eqforquadric_comp}
\frac{\p}{\p x^a} \left(\, b^{ab}(u) \, \frac{\p u}{\p x^b}  \, \right) = 0,
\ee
where ${\bf b}(u)$ is a symmetric $N\times N$ matrix.
While the ansatz can be traced back to some works of Darboux \cite{Darboux},
in the `modern' times it has been applied to a class of dispersionless,
as well as linear PDEs: the $SU(\infty)$--Toda equation \cite{T95},
the dKP equation \cite{DT02}, the Laplace equation \cite{D03}, as well
as a general class of equations integrable by the method of hydrodynamic reductions \cite{F12}. The idea is to find and  solve
an ODE for ${\bf M}(u)$ (and keep in mind that this ODE is not a
symmetry reduction of the underlying PDE, 
and the corresponding solutions in general do not admit any Lie--point, or generalised symmetries). The ODE arises as follows:
Differentiating (\ref{quadricansatz}) 
implicitly w.r.t. $x^a$
and substituting the resulting expression for $\p u/\p x^a$ into 
(\ref{eqforquadric_comp}) yields a matrix ODE
\[
g {\bf M}' = {\bf M b M},\quad\mbox{where}\quad'=\frac{d}{du}
\]
and the function $g=g(u)$ is defined by $2g'=\mbox{Tr}({\bf bM})$.
Setting ${\bf N} = - {\bf M}^{-1},$ this matrix ODE simplifies to 
\be \label{Neq} g {\bf N}' = {\bf b},
\ee
and one finds that
\be \label{geq} \quad  g^2 \det {\bf N} =  \zeta, \ee
where $\zeta$ is an arbitrary constant.

From now, we shall regard  the system (\ref{Neq}, \ref{geq}) as the reduction of (\ref{eqforquadric_comp}) 
under the quadric ansatz  (\ref{quadricansatz}).   Solving this system  for the components of ${\bf N}$ gives the matrix  ${\bf M},$ and thus
leads to an implicit solution to the nonlinear PDE (\ref{eqforquadric_comp}) of the form (\ref{quadricansatz}), which is therefore constant on a central quadric.
\vskip5pt
The $mn$-dKP equation (\ref{mndKP}) is of the form (\ref{eqforquadric_comp}), with 
\be \label{mndKPmatrixb}    {\bf b}(u)   =
  \left(\begin{array}{c|ccc|c}
   -u^m & 0 & \cdots &  0 &  \; \frac{1}{2} \\ \hline
    0  &   & & &   \;  0 \\
\vdots & &{\scalebox{1.2}{$-{\bf 1}_n$}} &  & \; \vdots \\
    0 & & & &  \; 0  \\ \hline
\frac{1}{2} & 0 & \cdots &  0 & \;  0
  \end{array}\right),
\ee
where ${\bf 1}_n$ denotes the $n \times n$ identity matrix. 
Equation  (\ref{Neq}) then implies that
\be \label{matrixN}   {\bf N}   =
  \left(\begin{array}{c|ccc|c}
   Y & \beta_1 & \cdots &  \beta_n &  \; Z  \\ \hline
   \beta_1  &   & & &   \;  \vv_1 \\
\vdots & &{\scalebox{1.2}{${\bf X}$}} &  & \; \vdots \\
    \beta_n & & & &  \; \vv_n  \\ \hline
Z & \vv_1 & \cdots & \vv_n & \;  \phi
  \end{array}\right),
\ee
where $\beta_i, \vv_i,$ $i = 1,\dots, n,$ and $\phi$ are constants,  $Y$ and $Z$ are functions of $u,$  and ${\bf X}$ is an $n \times n$ symmetric matrix,
whose diagonal components $X_1, X_2, \dots, X_n$
are functions of $u$ and off-diagonal components $\alpha_{ij}, i>j$ are all constants.  
Equation  (\ref{Neq}) also implies that 
\[ 
Y' = -g^{-1}{u^m},\quad {X_i}' = -g^{-1}, \quad  Z' = (2 g)^{-1}.
\]
The last equation gives $g$ in terms of  $Z'$, so that
\be \label{mndKPXYZ}    
Y' + 2 u^m Z'  = 0, 
\ee
and
\be
\label{matrixX}
{\bf X}(u)=-2Z(u){\bf 1}_n+{\bf X}_0
\ee
where ${\bf X}_0$ is a constant matrix with diagonal components $\gamma_i$ and
off--diagonal components $\alpha_{ij}$.
The equation (\ref{geq}) becomes
\be \label{algY}  \det {\bf N} \; = \; 4 \, \zeta \, \left( Z' \right)^2, \ee
and, together with the form (\ref{matrixN}) gives an algebraic expression for $Y$ in terms of $Z, Z'$
and constants. Substituting this expression into (\ref{mndKPXYZ}) yields a second order scalar
ODE for $Z$ of the form
\be
\label{geneqcase2}
Z'' = A(Z) \, \left( Z' \right)^2 + B(Z) + C(Z) u^m, 
\ee
where $A, B$ and $C$ are rational functions of $Z$,
of degrees respectively $n, 2n+1$ and $n$, which will be determined in \S\ref{section_sol}.

 If this equation can be solved, then reversing the steps above we can reconstruct the matrix
${\bf M}(u)$, and thus find an explicit solution to (\ref{mndKP}) constant on a central quadric. All solutions
in the class (\ref{quadricansatz}) arise from this construction.
\subsection{Eliminating the constants}
To make further progress we exploit some symmetries of the $mn$--dKP equation (\ref{mndKP}) to 
eliminate some of the constants in (\ref{matrixN}). If $m=1$, then the transformation 
\be 
\label{symm1}
x^a\rightarrow \hat{x}^a={A^a}_b x^b, \quad\mbox{where}\quad
{\bf A} = \left( 
\begin{array}{cccccc}
1 & c_1 & c_2 & \dots & c_n & k \\
 & 1 &  0 &  \dots &  0 & 2 c_1 \\
&  & 1 &   \ddots & \vdots & 2 c_2 \\
 &  &  & \ddots &  0 &  \vdots \\
 &  {\scalebox{2}{$0$}} &   &   &  1 & 2 c_n \\
 &  &  &  &   & 1 
\end{array}  \right) 
\ee
together with
\be
u\rightarrow
\hat{u}  = u + c_1^2 + c_2^2 + \dots + c_n^2 - k,
\ee
where $c_i,$ $i = 1, \dots, n,$ and $k$ are constants is a symmetry of (\ref{mndKP}), 
which also preserves the quadric ansatz (\ref{quadricansatz}) with a replacement
\be
\label{hNN}
{\bf M}(u)\rightarrow{\bf{\hat{M}}}(\hat{u})=({\bf A}^{-1})^T
{\bf M}(u){\bf A}^{-1}, \quad\mbox{or}\quad 
{\bf \hat{N}}(\hat{u}) =   {\bf A} \,  {\bf N}(u) \, {\bf A}^T.
\ee
Using (\ref{hNN}) we can set some of the constants $(\varepsilon, \beta, \alpha, \gamma)$  to $0$.
There are three cases to consider depending on whether $\phi$ vanishes, or not.
\section{Solutions}
\label{section_sol}
\subsection{Case I}
\label{sec_examples}
Assuming $\phi=\varepsilon_i=0$ for $i=1, \dots, n$ leads to the quadric ansatz equations which are solvable by quadrature, 
\be \label{uY}
u =  2 \int \left(\frac{-\zeta}{Z^2 \det {\bf X}}\right)^{\frac{1}{2}} \; dZ,
  \quad Y = -2 \int u^m \, \frac{d Z}{du} \; du.
\ee
To derive these formulae note that in this case $\mbox{det}({\bf N})$ is independent on $Y$. Therefore (\ref{uY}) arises from 
\be
\label{my_new}
4\zeta(Z')^2=-Z^2\mbox{det}({\bf X})
\ee
which is
(\ref{algY}), where $\mbox{det}({\bf X})$ is a polynomial of degree $n$ in $Z$.
If $n=1$ or $2$, then the corresponding expression for $u$ is given by elementary functions.
Two examples of such solutions are given below:
\subsubsection {$(m, n)=(2, 1)$}
\begin{eqnarray}\frac{1}{\tan^2 u} \Bigg[\frac{y^2}{4}   &+& \Big( 2 \sin^2 u \cos u \, \big(\cos u \ln(\cos u) + u \sin u \big)  -  u^2 \sin^2 u + \delta^2 \cos^4 u  \Big) t^2   \nonumber \\
                                                         &-& \delta (\cos^2 u)  y t  - (\sin^2 u) x t \Bigg] \; = \;  C,  \label{solcase1n1m2}\quad\mbox{where}\quad\delta=\mbox{const.}
\end{eqnarray}
\subsubsection{$(m, n)=(1, 2)$}
\be \label{case1explicitsol}  (4 a e^{au} + 1) \Big[ 4 a^3 (y_1^2 + y_2^2) + 4 e^{-a u} \big( a^2 xt - at^2 \ln (e^{-au} + 4a) \big) - e^{-2au} t^2 \big( au + \ln(e^{-au} + 4a) \big) \Big]  = C,\ee  
where $a>0$ is a constant.
\subsubsection{$m=1$} There is also a simple class of solutions with $m=1$, and arbitrary 
$n$. To find it suppose that all constants in the matrices ${\bf N}$ and ${\bf X}_0$ are zero. 
Then equation (\ref{my_new})
with $\zeta = (-1)^{n+1} 2^{n-2}$ becomes $(Z')^2=Z^{n+2}$, which after integrating (\ref{uY})
gives
\begin{eqnarray*}
 u \, \Big( 4 u (\ln u)\, t^2  - 4xt + y_1^2 + y_2^2  \Big)   &=& C, \quad n=2, \\
u^{2/n} \left( \frac{8}{n-2} u\, t^2 - 4xt + \sum_{i=1}^n y_i^2 \right)  &=& C, \quad n \ne 2.
\end{eqnarray*}
\subsection{Case II}
We shall now assume that the constant $\phi$ in (\ref{matrixN}) is zero, but at least one of the $\varepsilon_i$s is non--zero. If $m=1$, the symmetry
(\ref{symm1}) can be used to eliminate all $\beta$s and $\gamma_1$. Then (\ref{algY}) 
with $R_{n}(Z)\equiv \mbox{det}({\bf X})$ gives
\be
\label{detNeqcase2} 4 \, \zeta \, \left( Z' \right)^2  \; = \;  Y Q_{n-1}(Z) - Z^2 R_{n}(Z),\quad\mbox{where}
\quad Q_{n-1}(Z)\equiv  \det 
  \left(\begin{array}{ccc|c}
      & & &   \;  \vv_1 \\
 &{\scalebox{1.2}{${\bf X}$}} &  & \; \vdots \\
     & & &  \; \vv_n  \\ \hline
 \vv_1 & \cdots & \vv_n & \;  0
  \end{array}\right).
\ee
Differentiating (\ref{detNeqcase2}) with respect to $u$ and substituting
$Y' = -2 u Z'$ gives
a second order ODE for $Z$ of the form (\ref{geneqcase2}) where $m=1$, and
\[ A(Z) =   \frac{1}{2}  \frac{d}{dZ} \left(\ln Q_{n-1}\right), \; 
B(Z) = \frac{1}{8 \zeta} \left(  Z^2 R_n \frac{d}{dZ} (\ln Q_{n-1})  - \frac{d}{dZ} (Z^2 R_n) \right), \; 
C(Z)= - \frac{ Q_{n-1}}{4 \zeta}. 
\]
\begin{prop}
\label{prop_p1}
Let $\phi=0$, and at least one of the $\varepsilon_i$s is non--zero in (\ref{matrixN}).
  If $(m, n)=(1, 1)$, the equation (\ref{geneqcase2}) is equivalent to Painlev\'e I. For $m=1$, and $n>1$
  equation (\ref{geneqcase2}) does not posses the Painlev\'e property.
\end{prop}\noindent
{\bf Proof.}
For an ODE to have the Painlev\'e property, its movable singularities  can only be poles.  Thus we follow 
the algorithm in \cite{ARS80} to determine whether the general solution of (\ref{geneqcase2}) admits a movable branch point.  
For convenience we first conduct the Painlev\'e test in the special case, where the constant matrix ${\bf X}_0$ is zero, 
so that ${\bf X}=-2Z{\bf 1}_n$, and (with the definition
$\varepsilon^2\equiv {\varepsilon_1}^2+\dots+ {\varepsilon_n}^2$)
\be \label{geneqcase2simp}  
2 \zeta \, Z'' = \zeta \frac{(n-1)}{Z} \left(Z'\right)^2  - 3 (-2)^{n-2}  Z^{n+1} - \vv^2 (-2)^{n-2}  Z^{n-1} u.
\ee
Assume that the dominant behaviour of a solution near a movable singularity $u_0$ is of the form 
\be \label{dominantap} Z \approx a (u-u_0)^p,\ee
where $a$ and $p$ are constants.
Then substitute (\ref{dominantap}) into (\ref{geneqcase2simp}) and balance the power of $u-u_0$ of two or more terms.  If the
balancing terms are dominant, i.e. their power of $u-u_0$ is most negative, then other terms can be ignored, and one can solve for $a.$  
For equation (\ref{geneqcase2simp}), it turns out that the only possible value of $p$ is $ p = -\frac{2}{n},$ which is not an integer  for $n>2$ and  suggests 
a movable algebraic branch point.  Moreover, this result 
extends to the general case  (\ref{geneqcase2}).  This is because the assumption that leads to the special case (\ref{geneqcase2simp}) 
keeps only the highest degree terms in  $Q_{n-1}$ and  $R_{n}.$  The presence of the lower degree terms in the rational functions $A(Z),$ $B(Z)$ 
and $C(Z)$ will not change the possible dominant behaviour in the first step of the Painlev\'e test.  Therefore we  conclude that equation 
(\ref{geneqcase2}) does not posses the Painlev\'e property for $n>2.$ 

If $n=1,$ then (\ref{geneqcase2simp}) and
(\ref{geneqcase2}) are equivalent, after constant rescalings of dependent and independent variables, to the Painlev\'e I equation. 

 If $n=2,$ with $\beta$s and $\gamma_1$ eliminated by the symmetry for $m=1,$ then (\ref{geneqcase2}) becomes
\be \label{geneqcase2n2}  
4 \zeta \, Z'' = \frac{4\zeta}{2Z + \delta} \left(Z'\right)^2  +
\frac{4 Z^4 - 2 \g Z^3 - \a^2 Z^2}{2Z + \delta} - 8  Z^{3}  + 3\g Z^2 + \a^2 Z 
   - \vv^2 (2Z + \delta) u, \ee
where we let $\vv^2 \equiv \vv_1^2 + \vv_2^2,$ $\g\equiv \g_2$ and $\delta \equiv \frac{2 \a \vv_1 \vv_2 - \g \vv_1^2}{\vv_1^2 + \vv_2^2}.$ 
Here $\a$ is the off--diagonal component of
the matrix ${\bf X}_0.$
  Substituting  (\ref{dominantap}) in (\ref{geneqcase2n2}), 
the only possibility is $(p,a) = (-1, (-\zeta)^{1/2}).$  
Let $- \zeta = \kappa^2,$ and take $a = \kappa.$  The remaining steps in the algorithm will determine whether 
the general solution of (\ref{geneqcase2n2}) can be represented near a movable singular point $u_0$ by the Laurent series, with the leading term
$ \frac{\kappa}{(u-u_0)}.$ \, 
It turns out that to satisfy (\ref{geneqcase2n2})  one needs to introduce a logarithmic term, which gives 
\[ Z(u) \approx \frac{\kappa}{(u-u_0)} + \frac{\g}{8} - \frac{32 \vv^2 u_0 - 3 \g^2 - 8 \a^2}{192\kappa} (u-u_0)  + \left(c  + \frac{\vv^2}{8\kappa} \ln(u-u_0) \right) (u-u_0)^2 
+ O ( (u-u_0)^2), \]
where $c$ is an arbitrary constant. The logarithmic term indicates a  logarithmic branch point, and this shows that (\ref{geneqcase2n2})  does not
posses the Painlev\'e property. Hence we  conclude that equation 
(\ref{geneqcase2})  does not posses the Painlev\'e property for $m=1,$ $n>1.$
\koniec
\subsection{Case III}
This is the generic case, where we assume that $\phi\neq 0 $ in
(\ref{matrixN}). If ${m=n=1}$, then the ODE resulting from the quadric anzats  reduces to
Painlev\'e II \cite{DT02}. For general $n$, and $m=1$ the symmetry (\ref{symm1})
can be used to eliminate $\gamma_1$ and all $\varepsilon$s. Equation (\ref{algY})
takes the form
\be \label{detNeqcase3} 4 \, \zeta \, \left( Z' \right)^2  \; = \;    (\phi Y -  Z^2) R_n(Z) + Q_{n-1}(Z), \ee
where $R_{n}(Z) = \det {\bf X}$, and here  $Q_{n-1}(Z) = - \phi \sum_{i=1}^n \beta_i \, \det {\bf B}_i$ 
with ${\bf B}_i$ denoting the matrices obtained from replacing the $i$th column of ${\bf X}$ (\ref{matrixX}) by the column vector $(\beta_1, \dots, \beta_n)^T.$
Solving this for $Y$ in terms of $Z$ and $Z'$, and differentiating to eliminate
$Y'$ by $Y'=-2u Z'$ gives
(\ref{geneqcase2})
where $m=1$, and
\[ A(Z) =   \frac{1}{2}  \frac{d}{dZ} \left(\ln R_{n}\right), \; 
B(Z) = \frac{1}{8 \zeta} \left(\frac{d Q_{n-1}}{dZ}  - Q_{n-1}\frac{d}{dZ} (\ln R_{n})  -2 Z R_n\right), \; 
C(Z) = - \frac{\phi}{4 \zeta}  R_{n}. \]
\begin{prop}
\label{prop_p2}
Let $\phi\neq 0$ in (\ref{matrixN}).
If $(m, n)=(1, 1)$, the equation (\ref{geneqcase2}) is equivalent to Painlev\'e II. For $m=1$ and $n>1$
equation (\ref{geneqcase2}) does not posses the Painlev\'e property.
\end{prop}
\noindent
{\bf Proof.} The result is obtained by first performing the  Painlev\'e test on the 
special case  case where ${{\bf X} = -2Z  \,{\bf 1}_n}$ (i.e. assuming  ${\bf X}_0=0$ in (\ref{matrixX})) 
and (\ref{geneqcase2}) is
\be \label{geneqcase3simp}  
2 \zeta \, Z'' =  \frac{n \zeta}{Z} \left(Z'\right)^2  + (-2)^{n-1}  \left( Z^{n+1} + \frac{\phi \beta^2}{4} Z^{n-2} \right) 
+  (-2)^{n-1} \phi  Z^{n} u,
\ee
where  $\beta^2\equiv\beta_1^2 + \dots +\beta_n^2.$   When $n=1$, then, after a coordinate transformation
\cite{DT02} this family of ODEs (\ref{geneqcase3simp}) gives the Painlev\'e II equation. For $n>2$, the ODE (\ref{geneqcase3simp}) fails the test at the first step of finding the dominant behaviour of the general solution, where
it displays the dominant term of the form $a (u-u_0)^p$ with  $ p = -\frac{2}{n}.$  
\, For $n>2,$ this indicates an algebraic branch point of order $ -\frac{2}{n},$ hence (\ref{geneqcase3simp}) does not have the  Painlev\'e property.  Then 
we argue that this result extends to the general form  (\ref{geneqcase2}) as   (\ref{geneqcase2}) differs from
 (\ref{geneqcase3simp}) only by
 the lower degree terms in the polynomials appearing in the rational functions $A(Z), B(Z)$ and $C(Z),$ and these will not affect the dominant behaviour analysis.

 For $n=2,$
the form of (\ref{geneqcase2}) is still quite complicated by the presence  of the constants $\a$ and $\g=\gamma_2$ in the matrix ${\bf X}$.  
After a translational change of variable $Z \to \hat{Z} = Z -\g/4,$ and then dropping the hat, (\ref{geneqcase2}) becomes
\be \label{geneqcase3n2}  
 \zeta \, Z'' = \frac{\zeta Z}{Z^2  - \rho^2} \left(Z'\right)^2  - (Z+\frac{\g}{4}) (Z^2  - \rho^2)  - \frac{\phi \beta^2}{4} (2Z+\delta)  \frac{ Z }{Z^2  - \rho^2}  +
\frac{\phi \beta^2}{4}  - \phi (Z^2  - \rho^2) u, 
\ee
where $\beta^2 = \beta_1^2 + \beta_2^2,$ \,  $\delta = \frac{4 \a \beta_1 \beta_2 + {\g} (\beta_2^2 - \beta_1^2)}{2\beta^2}$ and 
$\rho^2 = \dfrac{4 \a^2 +\g^2}{16}.$  

\vspace{0.3cm}

The Painlev\'e test then shows that the general solution of (\ref{geneqcase3n2}) has a logarithmic branch point 
\[ Z(u) \approx \frac{\kappa}{(u-u_0)} - \frac{\phi u_0}{2}+ \frac{\g}{8}  + \left(c  + \frac{\phi}{3} \ln(u-u_0) \right) (u-u_0) 
+ O ( u-u_0), \]
where $\kappa = \sqrt{-\zeta}$ and $c$ and $u_0$ are arbitrary constants.  Hence we conclude that  (\ref{geneqcase2}) does not have the 
Painlev\'e property for $m=1,$ $n=2$.
\koniec
If $m>1,$ there is no obvious symmetry to eliminate the constants in ${\bf N}.$  Nevertheless, the Painlev\'e analysis shows that the case $m=n=1$ is the only case that the quadric ansatz reduction possesses the Painlev\'e property.

\begin{theo}
\label{theo1}
The quadric ansatz reduction  of the $mn$-dKP equation does not posses the Painlev\'e property unless ${m=n=1.}$
\end{theo}
\noindent {\bf Proof.}  The quadric ansatz reduction is of the form (\ref{geneqcase2}),
where $A(Z)$ and $B(Z)$ are rational functions of degrees respectively $n$ and $2n+1$, and $C(Z)$ is a polynomial of degree $n$.
The non--zero constants in ${\bf N}$ for $m \ge 1$ only contribute to the lower degree terms in the polynomials appearing in $(A(Z), B(Z), C(Z))$ and thus will not change the dominant behaviour of a solution near a movable singularity $u_0.$  Also,  the term $C(Z) u^m$ is not leading for any $m.$ Therefore, from the proofs of Propositions \ref{prop_p1} and \ref{prop_p2} we conclude that for $n>2,$ and any $m\geq 1$ the general solution has a movable algebraic branch point of order  $-\frac{2}{n}.$

It remains to settle the case where $n=1$ or $2$, and $m>1$. Performing the Painlev\'e test in these cases we find that the  general solution exhibits a logarithmic branch point.  In particular,  we have the following form of the general solution:

\noindent {\bf Case II.} \,  $\phi=0$ and at least one of the $\vv_i$s is non--zero in (\ref{matrixN}).

\noindent  $n=1:$
\vspace{-0.5cm}
\begin{eqnarray*} Z(u) \; \approx  &{}& \frac{8 \zeta}{(u-u_0)^2} \, + \, \frac{\g}{6} \, - \, \frac{12 \vv (u_0^m \vv +\beta) - \g^2}{480\zeta} (u-u_0)^2 \,  - \,  \frac{m u_0^{m-1} \vv^2}{24 \zeta} (u-u_0)^3 \\
&{}& + \, \left(c  + \frac{m(m-1) u_0^{m-2}\vv^2}{56\zeta} \ln(u-u_0) \right) (u-u_0)^4 \, + \, O ( (u-u_0)^4)
\end{eqnarray*}

\noindent  $n=2:$
\vspace{-0.5cm}
\begin{eqnarray*}Z(u) \; \approx &{}& \frac{\kappa}{(u-u_0)} \, + \, \frac{\g_1 + \g_2}{8} \, - \, \frac{32 ( u_0^m(\vv_1^2 + \vv_2^2) +\beta_1 \vv_1 + \beta_2 \vv_2) - 8 \a^2 - 3 (\g_1^2+\g_2^2) + 2 \g_1 \g_2}{192\kappa} (u-u_0) \\
&{}&   + \,  \left(c  + \frac{m u_0^{m-1}(\vv_1^2 + \vv_2^2)}{8\kappa} \ln(u-u_0) \right) (u-u_0)^2  \,+ \,O ( (u-u_0)^2) 
\end{eqnarray*}

\noindent {\bf Case III.} \,  $\phi \ne 0$  in (\ref{matrixN}).

\noindent  $n=1:$
\vspace{-0.5cm}
\begin{eqnarray*}Z(u) \; \approx &{}& \frac{8 \zeta}{(u-u_0)^2} \, - \, \frac{2}{3} u_0^m \phi  \, +\, \frac{\g}{6} \, - \, m  u_0^{m-1} \phi  (u-u_0) \\
&{}& +  \left(c  + \frac{2m (m-1)  u_0^{m-2} \phi }{5} \ln(u-u_0) \right) (u-u_0)^2  \,+ \,O ( (u-u_0)^2) 
\end{eqnarray*}

\noindent  $n=2:$
\vspace{-0.5cm}
\[ Z(u) \approx \frac{\kappa}{(u-u_0)} - \frac{u_0^m \phi }{2}+ \frac{\g_1+\g_2}{8}   + \left(c  + \frac{m u_0^{m-1} \phi}{3} \ln(u-u_0) \right) (u-u_0) 
+ O ( u-u_0), \]
\noindent Here, $\kappa = \sqrt{-\zeta},$  $c$ is an arbitrary constant, and $\g \equiv \g_1$ for $n=1.$
\koniec

\section{Einstein--Weyl geometry}
\label{sec_ew}
There is a, by now well established, link between 2+1 dimensional Einstein--Weyl geometry, and dispersionless integrable systems \cite{Wtoda, D01, D_hydro, C, DK, KF}.
In particular, the Manakov--Santini equation \cite{MS1} is known to be the general local normal form of the
Einstein--Weyl equations \cite{DFK}.

In this section we construct a Weyl structure in an arbitrary
dimension $N=n+2$, such that the Einstein--Weyl condition reduces to a single
dispersionless PDE. In the case when $n=1$ this PDE is the dKP equation, and
the Einstein--Weyl structure is that of \cite{D01}, and for $n>1$
the PDE is (\ref{EWeqmndKP}). The main result (Theorem \ref{theo_EW}) is an explicit class of Einstein--Weyl
spaces depending on one arbitrary function of one variable.

Recall \cite{PT93} that a Weyl structure on a manifold $U$ (which is really just an open set in $\R^{N}$, as our considerations are local) consists of a conformal structure $[h]$
represented by a metric $h$, and a torsion--free connection $D$ which
is compatible with $[h]$ in the sense that $Dh=\nu\otimes h$ for some
one--form $\nu$. This compatibility is invariant under the conformal
change of metric:
\be
\label{conf_change}
h\rightarrow\Omega^2 h, \quad \nu\rightarrow \nu+2d(\ln{\Omega}),
\ee
where $\Omega:U\rightarrow\R^{+}$. A Weyl structure is said to be non--closed if $d\nu\neq 0$, or equivalently
if $D$ is not a Levi--Civita connection of any metric in the class $[h]$.
The Einstein–Weyl (EW) equations hold if the symmetrised Ricci tensor of $D$ is
proportional to some metric $h \in [h]$. The EW equations can be regarded as a system of PDEs
for the representative metric $h$, and the associated one--form $\nu$:
\be
\label{EWeq}
\chi_{ab} \equiv R_{ab} + \frac{N-2}{2} \, \nabla_{(a} \nu_{b)}    +   \frac{N-2}{4} \, \nu_a \nu_b 
- \frac{1}{N} \, h_{ab} \left( R + \frac{N-2}{2} \, \nabla_c \nu^c + \frac{N-2}{4} \, \nu_c \nu^c  \right) = 0,
\ee
where $\nabla, R_{ab}$ and $R$ are respectively the Levi--Civita connection, the Ricci tensor, and the Ricci scalar of $h$. A tensor $V$ is said to be of weight $k$, if $V\rightarrow \Omega^k V$ under the conformal rescaling
(\ref{conf_change}). If $V$ is a vector field of weight $k$, then the weighted covariant derivative of $V$ with respect to the Weyl connection is given by
\be
\label{weighted_der}
\widetilde{D}_a V^b=\nabla_a V^b-\frac{1}{2}\nu_c V^c {\delta_a}^b-\frac{1}{2}(k+1)\nu_a V^b+\frac{1}{2}V_a\nu^b.
\ee

The $mn$--dKP equation (\ref{mndKP}) can be written in the form
$d\star d u=0$, where the Hodge endomorphism $\star:\Lambda^{1}\rightarrow
\Lambda^{n+1}$  corresponds to the metric (note that the inverse metric corresponds to the matrix ${\bf b}(u)$ in
(\ref{eqforquadric_comp}))
\be
\label{metrich1}
h = dy_1^2 + \dots +dy_n^2-4 dx dt - 4 u^m dt^2.
\ee
In the case $m=n=1$ there exists a one--form $\nu=-4u_xdt$ such that the Einstein--Weyl condition reduces to the dKP equation \cite{D01}. It turns out that there is no
one--form which, together with the metric (\ref{metrich1}) gives the $mn$--dKP equation if $n>1$. We shall instead take the metric
(\ref{metrich1}) with $m=1$ (which can always be achieved by 
re-defining the function $u$) as a starting point. It can then be verified by an explicit computation
of (\ref{EWeq}) that for the
 Weyl structure represented by
\be \label{hmndKP}
h = dy_1^2 + \dots + dy_n^2 - 4 dx dt - 4 u dt^2, \quad \nu=-\frac{4}{n}u_x dt
\ee
with $x^0 :=t,$  $x^i :=y_i,$  $i=1, \dots, n,$ and $x^{n+1}:=x$, all components of $\chi_{ab}$ except $\chi_{00}$
vanish identically. The resulting Einstein--Weyl equation $\chi_{00}=0$ is a scalar PDE
(\ref{EWeqmndKP})
\[
u_{xt} - (u u_x)_x + \frac{2(n-1)}{n}u_x^2 = \triangle u.
\]
Moreover,  the vector field $V=\p/\p x$ is null, and covariantly constant with weight $-\frac{n}{2}$
with respect to $D$. Equation (\ref{EWeqmndKP}) is the dKP equation if $n=1$,
or its generalisation \cite{MOP, DG21} if $n>1$.

This class of Einstein--Weyl structures falls into a larger class of solutions which admit a parallel weighted
spinor \cite{DG21}. The particular case (\ref{hmndKP}) corresponds to Example 4 in this reference.
To understand a coordinate invariant characterisation of (\ref{hmndKP}), assume that an $(n+2)$--dimensional Weyl space represented by a pair $(h, \nu)$ admits a  covariantly constant null vector field with weight $-n/2$. 
Then the one--form ${\bf V}=h(V, \cdot)$ dual to $V$ satisfies
\be
\label{VVc}
d {\bf V}=\frac{4-n}{4}\nu\wedge {\bf V}.
\ee
The Frobenius theorem implies that there exist functions $(v, t)$ on $U$ so that
${\bf V}=vdt$. We shall use $t$ as one of the local coordinates on $U$. The existence
of canonical (up to some freedom) remaining $(n+1)$ coordinates $(y_1, \dots, y_n, x)$ is also guaranteed by the
Frobenius theorem: the distribution ${\mathcal V}$ of null curves
is spanned by $V=\p/\p x$, and its integrable orthogonal complement
${\mathcal V}^{\perp}$ is spanned by $\{\p/\p x, \p/\p y_1, \dots, \p/\p y_n\}$.

If $n\neq 4$ (so that the $\mbox{dim}(U)\neq 6$), the
we can rescale the metric so that ${\bf V}=-2dt$ and
\[
  h=f^{ij} \, dy_i dy_j-4dxdt+2A^idy_idt-4udt^2, \quad \nu=bdt,
\]
where the functions $f^{ij}, A^i, b$ and $u$ at this stage depend on all coordinates.
Going back to (\ref{weighted_der}) with $V=\p/\p x$ and $k=-n/2$, and considering its symmetrised part shows
that $f^{ij}$ and $A^i$ do not depend on $x$. Moreover
a coordinate transformation $y_i\rightarrow \hat{y}_i(y_j, t)$ together with
a redefinition of $f^{ij}(y, t)$ and $u(x, y, t)$ can be used to set $A^i=0$. The parallel weighted condition
on $V$ also imples that $b=-(4/n) u_x$.

The final step reducing the functions $f^{ij}$ to the identity $n\times n$ matrix is achieved in \cite{DG21}
by considering  the connection induced by $D$ on the screen bundle (see \cite{thomas}) 
\be
\label{screenb}
{\mathcal S}\equiv{\mathcal V}^{\perp}/{\mathcal V}\subset TU,
\ee
and restricting its holonomy to
$\R\otimes\mbox{Id}$ (equivalently the $\mathfrak{so}(n)$ projection of the holonomy algebra of this connection is zero). Now the metric and the one--form are given by  (\ref{hmndKP}), and the Einstein--Weyl equations reduce
to\footnote{In the special case $n=4$, we start--off with the metric
(\ref{hmndKP}), and the general one--form $\nu$ and impose the weighted parallel conditon on $V=\p/\p x$ to reduce the one--form $\nu$ to (\ref{hmndKP}).} (\ref{EWeqmndKP}).
\vskip5pt
To this end we shall construct an explicit subclass of examples of (\ref{hmndKP}) and (\ref{EWeqmndKP}) under the additional assumption that the screen bundle
distribution ${\mathcal V}^{\perp}/{\mathcal V}$ generates an isometric action
of $\R^n$ or $T^n$ on the Einstein--Weyl space. This will be done by linearising (\ref{EWeqmndKP}) by a contact 
transformation.
\begin{theo}
\label{theo_EW}
Let $(h, \nu)$ be an $(n+2)$--dimensional Einstein--Weyl structure $U$ which admits a parallel weighted null vector field $V$ with weight $-\frac{n}{2}$, and such that
\begin{itemize}
\item The connection on the screen bundle ${\mathcal S}$ defined by (\ref{screenb}) induced by $D$ has holonomy $\R\otimes{\mbox{Id}}$.
\item The sections of the screen bundle ${\mathcal S}$  generate the isometric action of the group of translations
$\R^n$ on $U$.
\end{itemize}
Then there exits local coordinates $(t, y_i, s)$ such that the one--form ${\bf V}\equiv h(V, \cdot)=-2dt$, the isometric action is generated
by $\{\p/\p y_1, \dots, \p/\p y_n\}$  and the Einstein--Weyl structure is given by
\begin{eqnarray}
\label{EWtheo1}
h&=&d{y_1}^2+\dots+d{y_n}^2+\frac{4G(s)}{(t-s)^{\frac{n}{n-2}}}dsdt, \quad \nu=-\frac{4}{(n-2)(t-s)}dt\quad\mbox{if $n\neq 2$}\nonumber\\
h&=&d{y_1}^2+d{y_2}^2+4G(s)e^{-st}dsdt, \quad \nu=-2sdt \quad\mbox{if $n=2$},
\end{eqnarray}
where $G=G(s)$ is an arbitrary function of one variable.
\end{theo}
\noindent
{\bf Proof.} 
The existence of the parallel weighed null vector field, and the holonomy reduction in Theorem \ref{theo_EW}
implies - as explained above - that the metric, and the one--form take the local normal form 
(\ref{hmndKP}). The additional symmetry assumption in the Theorem then implies that
$u=u(x, t)$, and the Einstein--Weyl condition (\ref{EWeqmndKP}) becomes
\be
\label{hh1}
u_{xt}-uu_{xx}+\kappa {u_x}^2=0, \quad\mbox{where}\quad \kappa=\frac{n-2}{n}.
\ee
Rewrite (\ref{hh1}) as a differential ideal
\begin{eqnarray}
\label{hh2}
\omega_1 &\equiv& du-u_xdx-u_tdt=0,\\
\omega_2 &\equiv& du_x\wedge dx+udu_x\wedge dt+\kappa {u_x}^2dt\wedge dx=0\nonumber
\end{eqnarray}
and set
\[
H=u-xu_x, \quad p=u_x
\]
Rewrtting  $\omega_1$ (\ref{hh2}) as
\begin{eqnarray*}
dH&=&u_tdt-xdu_x\\
&=&H_tdt+H_pdp
\end{eqnarray*}
gives
\be
\label{hh4}
x=-H_p,\quad u_t=H_t,\quad u=H-pH_p 
\ee
where now $H=H(p, t)$. Substituting (\ref{hh4}) into $\omega_2$ in (\ref{hh2}) gives
\be
\label{hh6}
H_{pt}+pH_p-H-\kappa p^2 H_{pp}=0.
\ee
The corresponding Einstein--Weyl structure (\ref{hmndKP}) takes the form
\be
\label{hh5}
h=d{y_1}^2+\dots+d{y_n}^2+4F(dpdt+\kappa p^2 dt^2), \quad \nu=-\frac{4}{n}pdt, \quad\mbox{where}\quad
F\equiv H_{pp}.
\ee
This only depends on the second derivatives of the function $H$, so the function $F$ is constrained by one
PDE obtained from differentiating (\ref{hh6}) with respect to $p$:
\be
F_t+(1-2k)pF-\kappa p^2 F_p=0.
\ee
This PDE can be solved explicitly, and the form of the general solution depends on $n=2/(1-k)$:
\begin{equation}
F=\begin{cases}
p^{-\frac{n-4}{n-2}} G\Big(t-\frac{n}{(n-2)p}\Big)& \text{if $n\neq 2$},\\
G(p)e^{-pt}, & \text{if $n=2$},
\end{cases}
\end{equation}
where in both cases $G$ is an arbitrary function of one variable. Introducing a new variable
$s$ by
\[
s=\begin{cases} t-\frac{n}{(n-2)p} & \text{if $n\neq 2$}, \\
p  & \text{if $n= 2$},
\end{cases}
\]
absorbing the overall constant into the arbitrary function $G$, and
adopting $(y_i, t, s)$ as local coordinates on $U$ yields (\ref{EWtheo1}).
\koniec 
\section{Conclusions}
We have demonstrated that solutions to the $mn$--dKP equation (\ref{mndKP}) constant on central quadrics are characterised by solutions to a 2nd order scalar ODE. In the generic case this ODE is of Painlev\'e type
if $m=n=1$, but does not posses the Painlev\'e property if $m\cdot n>1$.
This rules out the integrability of (\ref{mndKP}) for these values of $m, n$. There are other approaches to dispersionless integrability of (\ref{mndKP}) discussed in \cite{SS16}, and in particular
Boris Kruglikov informed us that the approach taken in references \cite{KF, CK} could also be used to rule out non--integrable cases, and perhaps narrow them down to $m=n=1$.

Equation (\ref{mndKP}) with $n=1$ has been studied numerically in \cite{DGK}, where an asymptotic description of a gradient catastrophe in generalised KP
equation was conjectured, and
related to special solutions of Painlev\'e I.
In \cite{SS16} the analytical approach to this shock formation has been presented. It would be interesting to understand whether our explicit solutions shed more light on these shock formations.

In \S\ref{sec_ew} we have related another multi--dimensional generalisation of the dKP equation 
(\ref{EWeqmndKP}) to a class of Lorentzian Einstein--Weyl structures. This class admits a parallel weighted null vector, and thus a parallel weighted 
spinor which makes it interesting in both physics (supersymmetric solutions to Einstein--Weyl equations \cite{MOP}), and geometry, where the existence of such spinor corresponds to a holonomy reduction of the Weyl connection \cite{DG21}.


\begin{thebibliography}{}

\bibitem{ARS80} Ablowitz, M. J.,  Ramani, A. and Segur, H. (1980) A connection between nonlinear evolution equations and ordinary differential 
equations of P-type. I, J. Math. Phys. {\bf 21}(4), 715-721.

\bibitem{C} Calderbank, D. M. J. (2014)
Integrable background geometries, SIGMA {\bf 10}, 034.

\bibitem{CK} Calderbank, D. M. J. and Kruglikov, B. (2021)
Integrability via Geometry: Dispersionless Differential Equations in Three and Four Dimensions. 
Comm. Math. Phys. {\bf 382}, 1811–1841.

\bibitem{Darboux} 
Darboux, G. (1910) Lecons sur les systmes orthogonaux et les coordonnes
curvilignes. Gauthiers-Villars, Paris.

\bibitem{DG21} Dikarev, A. and Galaev, A. S. (2021) Parallel spinors on Lorentzian Weyl spaces, Monatsh Math. https://doi.org/10.1007/s00605-021-01569-x.

\bibitem{DGK} Dubrovin, B., Grava, T. and Klein, C. (2016)
On critical behaviour in generalized Kadomtsev--Petviashvili equations. Nonlinearity {\bf 29},  
1384 - 1416.


\bibitem{D_hydro} Dunajski, M. (2009) {\em Solitons, Instantons and Twistors}. Oxford Graduate Texts in Mathematics {\bf 19}, OUP.

\bibitem{D03}  Dunajski, M.  (2003) Harmonic functions, central quadrics, and twistor theorey, Class. Quantum Grav.
{\bf 20}, 3427-3440. 

\bibitem{D01} Dunajski, M., Mason, L. J. and Tod, K. P. (2001) Einstein-Weyl geometry, the dKP equation and twistor theory, 
J. Geom. Phys. {\bf 37}, 63-92.


\bibitem{DT02} Dunajski, M. and Tod, K. P.  (2002) Einstein–Weyl spaces and dispersionless Kadomtsev–Petviashvili equation from Painlev\'e I and II,
Phys. Lett. A {\bf 303}(4), 253-264.

\bibitem{DK} Dunajski, M. and  Kry\'nski, W.
(2014)  Einstein--Weyl geometry, dispersionless Hirota equation and Veronese webs {\tt arXiv:1301.0621.} 
Math. Proc. Camb. Phil. Soc. {\bf 157}, 139-150


\bibitem{DFK} Dunajski, M.,  Ferapontov, E. and Kruglikov, B. (2015) 
On the Einstein-Weyl and conformal self-duality equations. {\tt arXiv:1406.0018}. Jour. Math. Phys. {\bf 56}


\bibitem{F12} Ferapontov, E. V., Huard, B. and Zhang, A. (2012) On the central quadric ansatz: integrable models and Painlev\'e reductions,  J. Phys. A: Math. Theor. {\bf 45},  195204.


\bibitem{KF} Ferapontov, E. and Kruglikov, B. (2014)  
Dispersionless integrable systems in 3D and Einstein-Weyl geometry
J. Differential Geom. {\bf 97}: 215-254. 

\bibitem{thomas} Leistner, T. (2006) Screen bundles of Lorentzian manifolds and some generalisations of 
$pp$--waves. J. Geom. Phys. {\bf 56}, 2117-2134.

\bibitem{MS1} Manakov, S. V. and Santini, P. M. (2006)
The Cauchy problem on the plane for the dispersionless Kadomtsev-Petviashvili equation, JETP Lett. {\bf 83}, 462–466.

\bibitem{MS11} Manakov, S. V. and Santini, P. M. (2011) On the dispersionless Kadomtsev–Petviashvili equation in $n+1$ dimensions:  exact solutions, the Cauchy problem for small initial data and wave breaking; J. Phys. A: Math. Theor. {\bf 44}, 405203.

\bibitem{MOP} Meessen, P, Ort\'in, T., and Palomo-Lozano, A. (2012)
On supersymmetric Einstein-Weyl spaces. J. Geom. Phys. {\bf 62} 301.

\bibitem{PT93} Pedersen, H. and Tod, K. P.  (1993) Three-dimensional Einstein-Weyl geometry, Adv. Math.
{\bf 97}, 74-109. 

\bibitem{SS16} Santucci, F. and Santini, P. M. (2016) On the dispersionless Kadomtsev–Petviashvili equation with arbitrary nonlinearity and dimensionality: exact solutions, longtime asymptotics of the Cauchy problem, wave breaking and shocks, J. Phys. A: Math. Theor. {\bf 49}, 405203.


\bibitem{T95}  Tod, K. P.  (1995) Scalar-flat K\"ahler and hyper-K\"ahler metrics from Painlev\'e-III, Class. Quantum Grav.
{\bf 12}, 1535-1547. 

\bibitem{Wtoda} Ward, R. S. (1990) Einstein-Weyl spaces and $SU(\infty)$ Toda fields, Classical Quantum 
Gravity {\bf 7}, L95-L98.

\end{thebibliography}
\end{document}